\documentclass[prl, twocolumn, amsmath,amssymb,superscriptaddress]{revtex4}

 \usepackage{setspace}

\usepackage{amsmath}
\usepackage{amsfonts}            
\usepackage{amssymb}    

\usepackage{mathrsfs}

\usepackage{eufrak}

\usepackage{graphicx}
\usepackage{epstopdf}

\usepackage{dcolumn}
\usepackage{bm}

\makeatletter
\def\@dotsep{4.5}
\makeatother

\newcommand*{\ket}[1]{\mathopen{|}#1\mathclose{\rangle}}

\begin{document}

\title{Dissipative binding of atoms by non-conservative forces}

\author{Mikhail Lemeshko}
\email{mlemeshko@cfa.harvard.edu}
\affiliation{ITAMP, Harvard-Smithsonian Center for Astrophysics, 60 Garden Street, Cambridge, MA 02138, USA}%
\affiliation{Physics Department, Harvard University, 17 Oxford Street, Cambridge, MA 02138, USA} %

\author{Hendrik Weimer}
\affiliation{ITAMP, Harvard-Smithsonian Center for Astrophysics, 60 Garden Street, Cambridge, MA 02138, USA}%
\affiliation{Physics Department, Harvard University, 17 Oxford Street, Cambridge, MA 02138, USA} %
\affiliation{Institut f\"ur Theoretische Physik, Leibniz Universit\"at Hannover, Appelstr. 2, 30167 Hannover, Germany}

 \begin{abstract}

  The formation of molecules and supramolecular structures results from bonding by conservative forces acting among electrons and nuclei and giving rise to equilibrium configurations defined by minima of the interaction potential. Here we show that bonding can also occur by the non-conservative forces responsible for interaction-induced  coherent population trapping. The bound state arises in a dissipative process and manifests itself as a stationary state at a preordained interatomic distance. Remarkably, such a dissipative bonding is present even when the interactions among the atoms are purely repulsive. The dissipative bound states can be created and studied spectroscopically in present-day experiments with ultracold atoms or molecules and can potentially serve for cooling strongly interacting quantum gases.

\end{abstract}

\date{\today}

\pacs{32.80.Qk, 34.20.Cf, 03.65.Yz, 37.10.De}

\maketitle

In most experiments investigating coherent quantum dynamics,
dissipation is an undesirable process. However, there have been several
recent theoretical proposals for turning controlled dissipation into a
useful resource, e.g., for the realization of interesting many-body
quantum states \cite{Diehl2008,Verstraete2009,Weimer2010,Diehl2010}. The main
idea is to engineer the interaction with the environment such that the
combination of coherent and dissipative dynamics drives the system
in question to a stationary state identical to the quantum state of interest. The feasibility of such a reservoir engineering has already
been demonstrated experimentally~\cite{Barreiro2011}.

In our work, we employ a similar idea to show that a bond between two interacting atoms (or molecules) can be induced by dissipation, thereby extending the notion of a bonding
mechanism from purely conservative to dissipative forces. The bond manifests itself as a stationary state  of the scattering dynamics that confines the atoms at a fixed distance. Owing to the
unprecedented control available for ultracold atoms and molecules~\cite{BlochRMP08}, the characteristics of the resulting molecules such
as bond lengths and spectroscopic properties are highly tunable and can be observed in current experiments.

\begin{figure}[t]
  \includegraphics[width=0.95\linewidth]{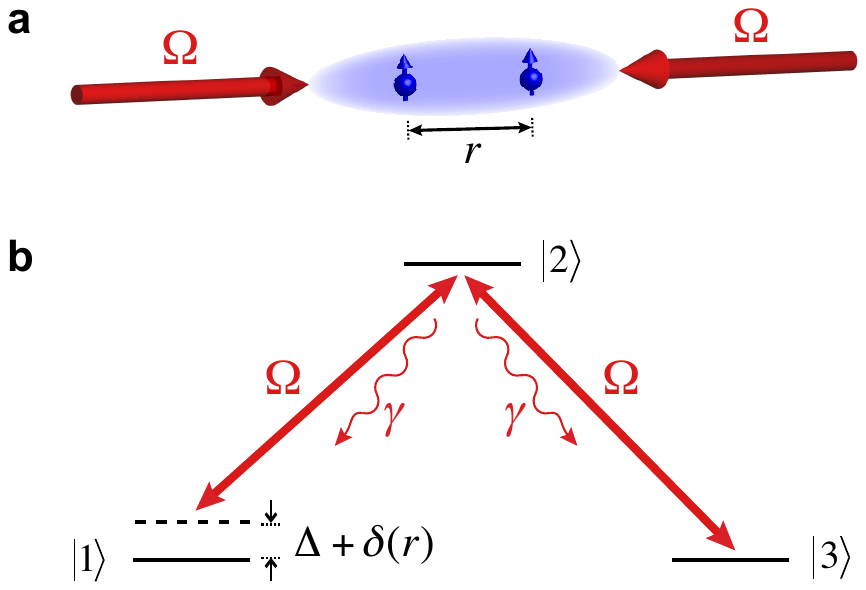}
  \caption{\label{fig:setup} \textbf{Setup of the system.} (\textbf{a}) Strong
    confinement by external fields restricts the atoms' motion to
    one spatial dimension parallel to the two counter-propagating
    laser beams represented by the driving fields with Rabi frequencies $\Omega$; (\textbf{b})~Internal level structure of the atoms. Two internal states
    $\ket{1}$ and $\ket{3}$ are coupled to a metastable state
    $\ket{2}$ spontaneously decaying with a rate $\gamma$. The laser field coupling the states $\ket{1}$ and $\ket{2}$ is detuned from the resonance by $\Delta$; state $\ket{1}$
    is subject to a dipolar interaction shift $\delta(r)$ which depends on the interparticle distance $r$; $\ket{2}$ and $\ket{3}$ are
    noninteracting.}
\end{figure}

As a specific physical implementation, we first consider two
interacting atoms whose motion is restricted to one spatial dimension
(1D), see Fig.~\ref{fig:setup}a. For ultracold
atoms and molecules, such a constraint can be enforced by appropriate
trapping potentials \cite{BlochRMP08}. We assume the internal level
structure of the atoms as shown in Fig.~\ref{fig:setup}b: each has
one internal state $\ket{1}$ exhibiting a distance-dependent
interaction shift, and two other noninteracting states $\ket{2}$ and
$\ket{3}$, with $\ket{2}$ undergoing spontaneous decay into states
$\ket{1}$ and $\ket{3}$. This particular level structure is quite
common in ultracold atoms and molecules, see the Methods section for
two different possible realizations based on Rydberg-dressed atoms
\cite{Henkel2010,Pupillo2010, Honer2010} or laser-cooled molecules
\cite{Stuhl2008,ShumanNature10,ManaiPRL12}. While our approach is
general, we focus on dipole-dipole interactions, as this brings in
additional tunability due to the anisotropy of the interaction
potential \cite{LahayePfauRPP2009}. We generalize our results to
higher spatial dimensions and many-particle systems, and discuss the
prospects of using dissipative bonding as a cooling mechanism for
strongly interacting quantum gases.

\begin{figure*}[t]
\includegraphics[width=\textwidth]{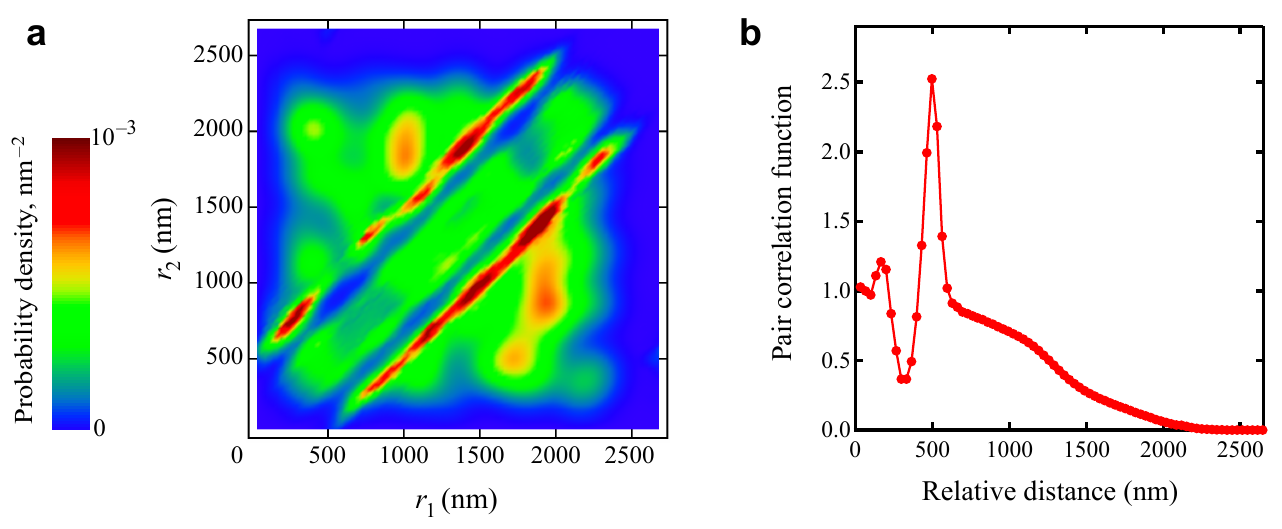}
\caption{\label{fig:distrib} \textbf{Formation of the dissipatively bound
  states.} (\textbf{a}) The probability density distribution shows that the
  absolute coordinates of the two particles, $r_1$, $r_2$, are
  delocalized; (\textbf{b}) the pair correlation function
  exhibits a sharp peak around $r_\text{d} =500\,\mathrm{nm}$.  Parameters
  correspond to a pair of Rydberg-dressed Cesium atoms (see the
  Methods section for details).}
\end{figure*}

\section{Results}

\subsection{Interaction-induced coherent population trapping}

The  process behind the formation of the bond is coherent
population trapping (CPT) \cite{Gray1978,Aspect1988}. In our setup, we
include two counter-propagating laser beams; the one is resonant with the transition between states $\ket{2}$ and
$\ket{3}$, and another is detuned by $\Delta$ from the resonance between $\ket{1}$ and $\ket{3}$; for simplicity we set the  Rabi frequencies of both fields to $\Omega$. CPT entails a dark state, i.e.,  a stationary state which cannot absorb photons. 

\begin{figure*}[t]
\includegraphics[width=\textwidth]{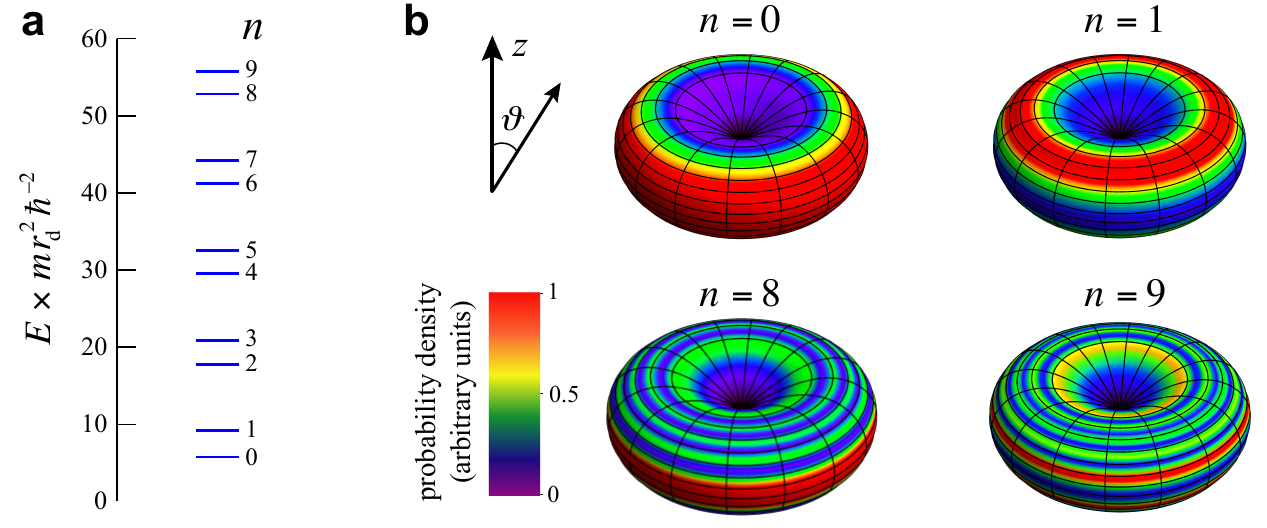}
\caption{\label{fig:donuts} \textbf{Three-dimensional eigenstates.}
  (\textbf{a}) Lowest energy levels featuring tunneling doublets due to the
  azimuthal symmetry of the interaction potential; (\textbf{b}) probability density
  distributions. The angle $\vartheta$ gives the  orientation of the collinear dipoles with respect to the vector connecting them. The quantum number $n$ gives the number of nodes of the probability density distribution (purple color) depending on the angle $\vartheta$. The results shown correspond to $J_z = 0$.}
\end{figure*}

For a system of two atoms subject to the electromagnetic field treated as a reservoir~\cite{Breuer2002}, the dynamics is obtained from the quantum master equation,
\begin{equation}
\label{eq:MasterEq}
	\frac{d \rho}{d t} = - i / \hbar \left[H, \rho \right] +
        \sum_{n} \gamma_{n} \left(c_{n}^{\phantom{\dagger}} \rho c_{n}^\dagger -
        \frac{1}{2} \{ c_{n}^\dagger c_{n}^{\phantom{\dagger}}, \rho \} \right),
\end{equation}
whose stationary solution is given by the condition $\mathrm{d} \rho /\mathrm{d}t  = 0$. Here,
$H$ is the Hamiltonian describing the two atoms and their
interaction with the laser fields, $\rho$ is the density operator
describing the two atoms, and $\gamma_n$ and $c_n$ are the rates
and corresponding jump operators associated with spontaneous decay,
see the Methods section for details. From the master equation~(\ref{eq:MasterEq}), one can extract an effective non-Hermitian Hamiltonian,
$H_\text{eff} = H - i V_\text{d}$, containing a dissipative potential,
\begin{equation}
\label{Vd}
V_\text{d} = \hbar \sum\limits_n \frac{\gamma_n}{2} c_n^\dagger c_n^{\phantom{\dagger}}.
\end{equation}
The zero-energy eigenstate of $H_\text{eff}$ corresponds to the dark state
$\ket{\psi_\mathrm{dark}}$ if its dissipation vanishes, i.e., $\langle
c_n^\dagger c_n^{\phantom{\dagger}}\rangle = 0$. In the
non-interacting but resonant case, $\delta(r) \equiv \Delta = 0$, the
dark state for each atom is given by
$(\ket{1}-\ket{3})/\sqrt{2}$. Taking the coupling to translation via the Doppler effect into account  retains this dark state and gives rise to the well-known velocity-selective CPT
\cite{Aspect1988}. In contrast, here we consider the regime where the kinetic energy is
small compared to the dipolar interaction between the atoms, which in turn is small compared to
both the laser driving and the decay, i.e., $(\hbar k)^2/2m \ll
\Delta, \delta(r) \ll \gamma,\Omega$.  We note that while the conventional   velocity-selective CPT scheme can preserve a dark state in presence of the dipole-dipole interactions between the ground and excited states~\cite{GoldsteinAPB95}, our setup involves interactions only in state $\vert 1 \rangle$ and therefore does not entail a true dark state corresponding to zero dissipation. In the limit of infinite mass
$m$, we find via perturbation theory, that the dissipation has a minimum
when the atoms are separated by a distance $r_\text{d}$ where the dipolar
interaction $\delta(r)$  destructively interferes with the detuning $\Delta$ the most,
\begin{equation}
\label{rd}
  r_\text{d} =  \left[\frac{d^2}{4\pi\varepsilon_0\hbar\Delta} \left(c_1+c_2\frac{\Omega^2}{\gamma^2}\right)  \right]^{1/3}.
\end{equation}
Here $d$ is the dipole moment of state $\ket{1}$, $\varepsilon_0$ the
vacuum permittivity, $\hbar$ Planck's constant, and the numerical
constants are $c_1 = (9+2\sqrt{2})/16$, and $c_2 =
(3+2\sqrt{2})/8$. The width of the Raman hole in the absorption
profile, on the order of $\Omega^2/\gamma$, amounts to
  the distances where the dipolar interaction and the detuning
  approximately cancel each other. While in conventional CPT schemes the photon scattering rate depends quadratically on the detuning from the resonance, the dipolar interaction shift scales with distance as $\delta(r) \sim 1/r^3$. This results in an anharmonic shape of the dissipative potential $V_\text{d}(r)$, and thereby in dependence of $r_\text{d}$ on the Rabi frequency $\Omega$ in Eq.~(\ref{rd}). Even though there still is some residual
dissipation for atoms confined to the vicinity of $r_\text{d}$, the
probability to find the atoms separated by $r_\text{d}$ is strongly
enhanced. Such a confinement of the atoms amounts to the formation of
a dissipative bond. Additional perturbations arising due to a finite
mass lead to a stationary state that exhibits a distribution of
distances sharply centered around $r_\text{d}$ rather than a single fixed
distance.

In the case of molecules bound by conservative forces, the strength of the bond is characterized by the binding energy -- the amount of energy needed to be transfered to a molecule for dissociation. Similarly, the strength of the dissipative bond can be characterized by an imaginary binding energy, reflecting the amount of \textit{dissipation} that is required to be added to the system in order to achieve dissociation. The imaginary binding energy is then defined as the difference between the decay rates at $r=\infty$ and $r=r_\text{d}$, and can be  calculated in the regime $\Omega  < \gamma$ as
\begin{equation}
\label{Binding}
E_\text{b} = i\hbar\gamma \frac{8 (\Delta/\gamma)^2}{ (4+9\sqrt{2}) (\Omega/\gamma) + (8+6\sqrt{2}) (\Omega/\gamma)^3}.
\end{equation}

\subsection{Numerical simulations of the dynamics}

We now turn to numerical simulations to confirm the existence of the
dissipative bond. We solve the quantum master
equation~(\ref{eq:MasterEq}) using the wave-function Monte-Carlo
method \cite{DalibardPRL92}, see the Methods section for details.
During the time evolution the atomic positions are redistributed by
the kinetic energy and photon recoil, and the atomic population is
accumulated in the vicinity of $r_\text{d}$ on a fast timescale of a few tens
of $\gamma^{-1}$. For an ensemble of atoms with randomly distributed
initial conditions we observe the appearance of a quasi-stationary
distribution strongly peaked around $r_\text{d}$, see Fig.~\ref{fig:distrib}.
For even longer times the spatial distribution of the atoms appears to
evolve towards a completely delocalized distribution of the laboratory-frame positions, $r_1$ and $r_2$,
compatible with a relative distance $r_\text{d}\approx 500\,\mathrm{nm}$.
This constitutes evidence that in the long-time limit the state
corresponding to the dissipative bond is a state with a relatively low
entropy, which is confirmed by the semiclassical studies of the
dynamics described below.  Once the dissipative bond has been formed,
it is possible to further reduce the probability of subsequent
spontaneous decay events by decreasing the Rabi frequencies
$\Omega$. This leads to an effective decay rate $\gamma_\text{eff}
=\gamma p_2 $, with $p_2$ the total probability of any of the atoms
being in state $\ket{2}$. For experimentally realistic parameters, see
the Methods section for details, we find $p_2 \approx 0.01$, implying
the lifetime of the dissipative bond to be longer than the timescale
of the formation of the bond.  Furthermore, we find that the reduction
of the Rabi frequencies does not lead to a significant increase of the
spread about $r_\text{d}$, and that lifetimes of the bound state exceeding
$\gamma_\text{eff}^{-1} \approx 0.1\,\mathrm{s}$ can be achieved.

\begin{figure*}[t]
\includegraphics[width=1.02\textwidth]{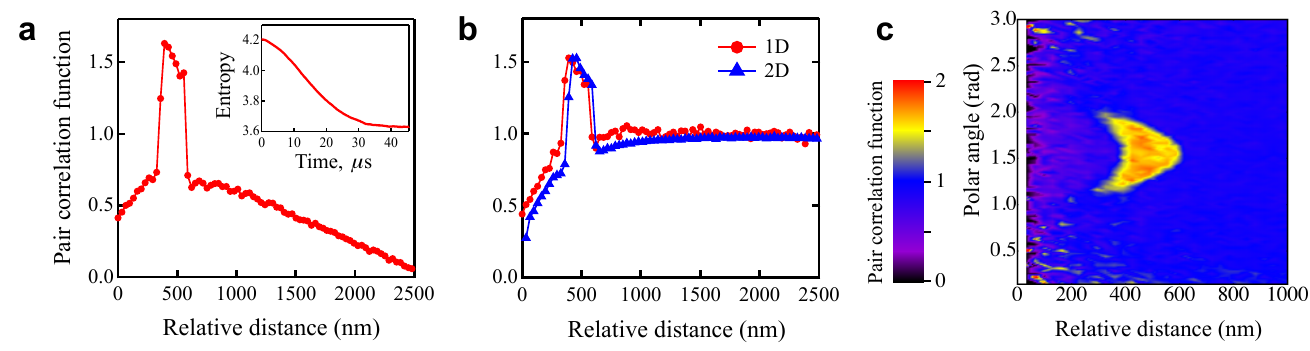}
\caption{\label{fig:semicl} \textbf{Semiclassical simulation of the
  dynamics.} (\textbf{a})  Pair correlation of two particles in
    1D, using the same parameters as in Fig.~\ref{fig:distrib}. The
  inset shows the time evolution of the relative
 distance entropy. (\textbf{b}) Many-body
    systems containing $10^2$ particles in 1D (red circles) and
  $10^4$ particles in 2D (blue triangles). (\textbf{c}) $10^4$ particles in 3D,
  with the polar angle $\vartheta$ as shown in
  Fig.~\ref{fig:donuts}b.}
\end{figure*}

The above results can be generalized to higher spatial dimensions,
thereby allowing to study the motion of the bound atoms. This merely
requires the inclusion of additional counter-propagating laser beams
as used in conventional laser cooling
setups~\cite{MetcalfLaserCooling}. For a two-dimensional
(2D) configuration, we consider the case when the
quantization axis defined by the electric dipole moment is
perpendicular to the plane of the motion, i.e., when the dipolar
interaction is always repulsive. Then, the dissipative bond restricts
the relative radial motion to a ring where the atoms are separated by
$r_\text{d}$. The effective Hamiltonian $H_\mathrm{eff}$ for the remaining
angular motion is then described by a two-dimensional rigid rotor,
$H_\mathrm{eff} = B J_z^2$, with $J_z$ the projection of the angular
momentum on the quantization axis. The rotational constant $B =
\hbar^2/(m r_\text{d}^2)$ can be widely tuned by changing $r_\text{d}$; for typical
experimental realizations, we find that values of $B \sim
100-1000\,\mathrm{Hz}$ are realistic, i.e., by far exceeding the
effective decay rate $\gamma_\mathrm{eff}$. The rotational spectrum of
the bound atoms can be then probed with standard techniques of
molecular spectroscopy.

In three spatial dimensions (3D), the constrained
relative motion of the two atoms is even richer due to the anisotropy
of the dipolar interaction. The dynamics is confined to a surface
defined by
\begin{equation}
r^3 = r_\text{d}^3 (1 - 3\cos^2 \vartheta),
\end{equation}
with $\vartheta$   the angle between the collinear dipoles and the interatomic radius-vector. Due to the rotational symmetry about the quantization
axis, the projection of the angular momentum, $J_z$, is conserved, allowing to reduce the problem to an effective one-dimensional
problem. The effective potential corresponds to an
azimuthally-symmetric double well centered at $\vartheta = \pi/2$,
resulting in an almost equidistant spectrum consisting of tunneling
doublets, Fig.~\ref{fig:donuts}a, similar to those encountered in microwave spectra of
ammonia~\cite{TownesSchawlow} or molecules in strong laser
fields~\cite{FriHerPRL95}. Fig.~\ref{fig:donuts}b features the
probability density distributions on the three-dimensional surface.
Although the results shown in Fig.~\ref{fig:donuts} correspond to
$J_z=0$, energy spectra and wavefunctions for other values of $J_z$
exhibit similar features.

Experimental observation of the resulting bound states can be performed by making use of the techniques well-established in the area of
ultracold quantum gases. In particular, a suitable setting would be a
many-body system, in which the atoms are essentially uncorrelated
at the beginning, with an average interatomic distance much larger
than $r_\text{d}$, so that many-body effects could be neglected. Then, the
existence of the bound state will appear as a sharp peak in the
pair-correlation function, see Fig.~\ref{fig:distrib}b, which is
readily accessible through Bragg scattering \cite{Stamper-Kurn1999} or
noise correlation spectroscopy \cite{Altman2004,Folling2005}.

In order to analyze the behavior in a many-body setting in more
detail, we employ a semiclassical model of the dynamics, see the
Methods section for the details. To compare this semiclassical
approach with the quantum mechanical model, we calculated the pair
correlation function for the case of two particles in a
one-dimensional trap, see Fig.~\ref{fig:semicl}a. Despite being
obtained in different temperature regimes, there is good agreement
with the quantum mechanical treatment of
Fig.~\ref{fig:distrib}b. The observed decrease in
  entropy towards a stationary value, shown in the inset of Fig.~\ref{fig:semicl}b, is a sign of
  a substantial cooling of the relative motion in the system. As
shown in Fig.~\ref{fig:semicl}b and c, the appearance of a sharp
peak around $r_\text{d}$ corresponding to the formation of dissipative bonds
is independent of the number of particles and the dimensionality of
the system. Note that in all the cases described above the density is
low enough to neglect any effects beyond two-body interactions.

\section{Discussion}

We have established the existence of a dissipative binding mechanism
triggered by non-conservative forces found in interaction-induced
coherent population trapping. Our results are applicable to a large
class of atomic and molecular systems, while the properties of the
bound states are highly tunable. Finally, we note that the mechanism
underlying the formation of the dissipative bond can potentially be
used for cooling of strongly-interacting many-body systems. Clearly,
the appearance of a low entropy stationary state independent of the
initial conditions already amounts to a demonstration of
 a cooling of the relative motion. In the high-density
regime, the correlation length $r_\text{d}$ will be modified by an effective
coordination number, embodying the interactions of an individual atom
with its surroundings in the stationary state. Despite this
renormalization, the formation of dissipatively bound complexes
\cite{Ates2012} in free space or even the realization of dipolar
crystals \cite{Buchler2007,Astrakharchik2007} appear possible.

\section{Methods}

\subsection{Quantum dynamics}

The coherent part of the dynamics according to the quantum master
equation, Eq.~(\ref{eq:MasterEq}), is given by the two-atom
Hamiltonian,
\begin{multline}
\label{Hamil}
	H = \sum_{k, i} \Biggl [ \frac{\hbar^2 k^2}{2 m} \vert k \rangle \langle k\vert_i - \frac{\Omega}{2} \left(\vert 1, k+\Delta k \rangle \langle 2, k \vert_i + \text{h.c.} \right) \\
	- \frac{\Omega}{2} \left(\vert 3, k-\Delta k \rangle \langle 2, k \vert_i  + \text{h.c.} \right)  - \Delta \vert 1, k \rangle \langle 1, k \vert_i  \Biggr] \\ 
	+ \sum_{k, k',q} \tilde{V}_{dd} (q) \vert 1, k-q\rangle_1 \vert 1, k'+q \rangle_2 \langle 1, k\vert_1\langle 1, k' \vert_2.
\end{multline}
Here, $i = 1,2$ and $k$ label the atoms and their corresponding
momentum states, and $\tilde{V}_{dd} (q)$ is the Fourier transform of the dipole-dipole
interaction potential. The dissipative part of Eq.~(\ref{eq:MasterEq})
contains the rates $\gamma_n = \gamma$ and jump operators $c_n =  \sum_k \vert
k+\Delta k_n, j_n \rangle \langle 2, k|_{i_n}$ in the Lindblad form,
responsible for the decay of each atom from state $\vert 2
\rangle$. The index $i_n=1,2$ runs over the two atoms, while
$j_n=1,3$ accounts for the two final states, and $\Delta k_n$ contains
all possible values of the emitted photon's wave vector \cite{DalibardPRL92}.

The results presented in Fig.~\ref{fig:distrib} correspond to a state after the evolution time of $4.6~\mu$s, averaged over 250 realizations with random initial conditions. In each realization two atoms are confined in a one-dimensional box $L\sim 6\,r_\text{d}$, which corresponds to the three-dimensional particle density of $3\times 10^{11}$~cm$^{-3}$. The initial state was chosen as a quasi-thermal distribution of Gaussian wavepackets having  an initial momentum of $4\,\hbar k_r$, where $k_r$ is the recoil momentum.

\subsection{Semiclassical dynamics}

For the semiclassical analysis, we use Langevin equations for the
motion of the particles. In analogy to the semiclassical description
of laser cooling \cite{Stenholm1986}, we introduce a velocity
dependent friction force and a stochastic heating force, which are
derived within perturbation theory. Additionally, our model includes a
transition rate to a bound channel, $\gamma_\text{b} = \Delta^2/(2 \sqrt{2}
\Omega)$, in which two bound particles move with the same
velocity. This binding term is only active for particles separated by
$|r-r_\text{d}|<w_\text{d}$, where $w_\text{d}$ is corresponds to the half-width at
half-maximum of the absorption profile. The reverse process, responsible for breaking of the dissipative bond, is governed by the stochastic heating force mentioned above. All results shown in Fig.~\ref{fig:semicl} were
obtained using the same atomic parameters and density as in the
quantum treatment, with the pair correlations computed at a time of
$4.6~\mu$s. As the semiclassical approach is only valid for momenta
much larger than $\hbar k_r$ \cite{Stenholm1986}, the initial momentum
distribution was chosen from a Boltzmann distribution with
$\sqrt{\langle k^2\rangle} = 40\,k_r$.  The relative distance entropy
was calculated as the Shannon entropy $S = -\sum_{r_i} p (r_i) \log
p(r_i)$, where $p (r_i)$ gives a probability to find atoms separated
by distance $r_i$ in the box.

\subsection{Experimental implementation}

We consider two different realizations of our setup depicted in
Fig.~\ref{fig:setup}, based on driven-dissipative Rydberg atoms
\cite{Lee2011,Glatzle2012,Honing2013} and laser-cooled molecules. In
the case of Rydberg-dressed Cesium atoms, the states $\vert 1 \rangle$
and $\vert 3 \rangle$ are chosen as different hyperfine components of
the electronic ground state, $6^2S_{1/2}$, of Cesium. The state $\vert
1 \rangle$ is provided with a dipole moment of $d = 15\,\mathrm{D}$
perpendicular to the 1D trap due to Rydberg dressing in an external
electric field~\cite{Henkel2010,Pupillo2010, Honer2010}; laser fields
$\Omega = \gamma/4$ drive the $6^2S_{1/2} \to 6^2P_{3/2}$ transition
having a linewidth of $\gamma = 2\pi \times 5.2$~MHz. These parameters
correspond to an imaginary binding energy of $E_\text{b} = i 2\pi\hbar \times
31$~kHz. We would like to stress that in contrast to the previous
Rydberg dressing proposals for the observation of interactions in a
Bose-Einstein condensate, the requirements on coherence times for the
observation of the dissipative bond are much less stringent;
preserving coherence on the timescale of the rotational constant
$B\sim 100\,\mathrm{Hz}$ can be achieved in present-day experiments
\cite{Schempp2010,Pritchard2010,Nipper2012,Dudin2012,Peyronel2012,
  SchaussNat12}, see Ref.~\cite{Low2012} for a recent review.

Although we exemplified our scheme by the case of bonding between a pair of atoms, it is equally applicable to bonding between a pair of molecules, such
as strontium monofluoride (SrF) that is being  laser-cooled \cite{ShumanNature10}. For
SrF, different hyperfine components, $F=0,1$, of the $X^2\Sigma^+
(v=0; N=1, J=1/2)$ state are chosen as $\vert 1 \rangle$ and $\vert 3
\rangle$. Here, a dipole moment of $3.5\,\mathrm{D}$ in the rotating
frame is imprinted on $\ket{1}$ via microwave
dressing~\cite{LemeshkoPRL12}, while the fields $\Omega =
\gamma/4$ drive transitions to the electronically excited
$A^2\Pi_{1/2} (v'=0; N=0, J'=1/2)$ whose natural linewidth is $\gamma
= 2\pi \times 7$~MHz. 

\subsection*{Acknowledgements}

We thank Bretislav Friedrich, Johannes Otterbach, Ignacio Cirac,  and Hossein Sadeghpour
for insightful discussions.
This work was supported by the National Science
Foundation through a grant for the Institute for
Theoretical Atomic, Molecular and Optical Physics at
Harvard University and Smithsonian Astrophysical
Observatory and within the Postdoc Program of the
German Academic Exchange Service (DAAD).

\subsection*{Author contributions} 

Both authors contributed equally to all parts of this work.

\subsection*{Additional information} 

\textbf{Competing financial interests.} The authors declare no competing financial interests.


\end{document}